\documentclass[journal=jpccck]{achemso}
\usepackage{amssymb}
\newcommand{\g}{graphene}

\SectionNumbersOn

    \author {Bhalchandra S. Pujari\footnote{Present Address: National Institute
    of Nanotechnology, 11421 Saskatchewan Drive,  Edmonton, Alberta, T6G 2M9, Canada.}}
    \email{bspujari@physics.unipune.ernet.in}
    \author {Dilip G. Kanhere}
    \affiliation{Department of Physics, University of Pune,  Ganeshkhind, Pune--411
    007, India.}
    \alsoaffiliation{Centre for Modelling and Simulations, University of Pune,
    Ganeshkhind, Pune--411 007, India.}
    \author {Biplab Sanyal}
    \affiliation{Division of Materials Theory, Department of Physics and Astronomy, Uppsala University, Box-516, SE 75120, Sweden}

    \title {Interaction of iron clusters (Fe$_n$; $n\le 6$) with a divacancy in graphene} 
                 
\begin{document}

\begin{abstract} 
    
In this work, we have studied the chemical and magnetic interactions of Fe$_n$;
$n\le 6$ clusters with a divacancy site in a graphene sheet by ab-initio density
functional calculations. Our results show  significant chemical interactions
between the cluster and graphene. As a result, a complex distribution of
magnetic moments appear on the distorted Fe clusters in presence of graphene and
results in a lower average magnetic moments compared to the free clusters.  The
presence of cluster also prevents the formation of 5-8-5 ringed structure known
to exhibit in  a graphene sheet having a divacancy defect.  The clusters induce
electronic states primarily of $d$-character near the Fermi level.

\end{abstract}

\section{Introduction}

Graphene has been a subject of immense investigation since its discovery in 2004
\cite{novoselov} as it has a great potential for future electronics \cite{geim,
netormp}.  It is a one-atom-thick planar sheet of sp$^2$-bonded carbon atoms
that are densely packed in a honeycomb crystal lattice.  Graphene is the basic
structural element of some carbon allotropes including graphite, carbon
nanotubes, fullerenes as well as recently synthesized graphane
\cite{elias01302009}.  Measurements have shown that graphene has a breaking
strength 200 times greater than steel, making it the strongest material ever
tested \cite{c:lee}.

Despite its strength, graphene is prone to the impurities and defects as any
other material.  The nature and types of defects in {\g} have been discussed by
Castro Neto {\it et al} in their extensive review. \cite{netormp} Among a number
of defects and disorders seen in {\g}, ripples or topological defects are
intrinsic while cracks, vacancies, charged impurities, atomic adsorption etc.
are extrinsic.  In particular {\g} is prone to the formation of vacancy defects.
Divacancy defects are quite probable to form as strong reactive centers. It is
shown theoretically \cite{lust} how different defect structures can be
engineered in graphene.  It is known that such defects cab affetct  the
electronic structure and hence transport properties of graphene \cite{coleman,
jafri, carva}.

One of the important points in graphene research is to explore the possibility
of having spin-dependent transport in graphene. As the mobility of graphene is
extraordinarily high, electronic transport with selective spin is an interesting
topic of study. The relevant question is how to make graphene magnetic.  One
idea is to deposit magnetic adatoms on graphene and study the range of spin
polarization in the host lattice arising from the exchange coupling between the
adatoms. One can make use of the defects for trapping the adatoms. It is known
that the chemisorption energies at vacancy sites are very high
\cite{biplabprb09}. So it is possible to trap the magnetic adatoms or clusters
at various defect sites and hope to have an effective spin polarization. In an
interesting theoretical work\cite{nieminen}  based on density functional theory
(DFT) the magnetism of different transition metal adatoms on single and
divacancy centers has been studied. In a work based on DFT, Wang {\it et al}
\cite{wang} have studied the interaction between a single adatom and graphene
containing a Stone-Wales defect. They have observed a reduction in local
magnetic moment on iron atom and a substantial modulation of electronic states
near the Fermi level.    Instead of a single adatom, a flux of adatoms may
generate various sizes of magnetic nanoclusters trapped in defect sites.  As
already noted, the divacancies are quite frequently form on a graphene sheet and
hence we investigate the interaction of small clusters of Fe with a divacancy.
Thus, a comparison of geometric and magnetic structure between adsorbed and free
Fe clusters will be made.

\section{Computational details}

All the calculations have been performed on a monolayer graphene with a
divacancy using plane-wave based density functional code VASP.\cite{vasp}  The
generalized gradient approximation as given by Perdew, Burke and Ernzerhof
\cite{PBE,PBEerr} has been used  for the exchange-correlation potential. The
energy and the Hellman-Feynman force thresholds are kept at 10$^{-5}$ eV and
10$^{-3}$ eV/{\AA} respectively. For geometry optimization, a 4 $\times$ 4
$\times$ 1 Monkhorst-Pack $k$-grid is used. Total energies and electronic
structures are calculated with the optimized structures on 11 $\times$ 11
$\times$ 1  Monkhorst-Pack $k$-grid. 

The supercell is generated by repeating the primitive cell by six times in $a$
and $b$ direction of the cell and then removing two carbon atoms in the vicinity
to create the divacancy. Such large cell, containing 70 atoms is required to
avoid the interaction between the clusters. Further interactions in vertical
direction is avoided by taking a vacuum of more than 15 {\AA}. 

To determine the ground state geometries, several initial structures of iron
clusters are placed on divacancy and are fully optimized. We wish to point out
that a special care has to be taken to achieve the correct magnetic moment by
starting the optimization procedure with several possible guesses of magnetic
states.

\section{Results and discussion}

\subsection{Geometries}

\begin{figure*}
\begin{center}
\includegraphics[width=14cm]{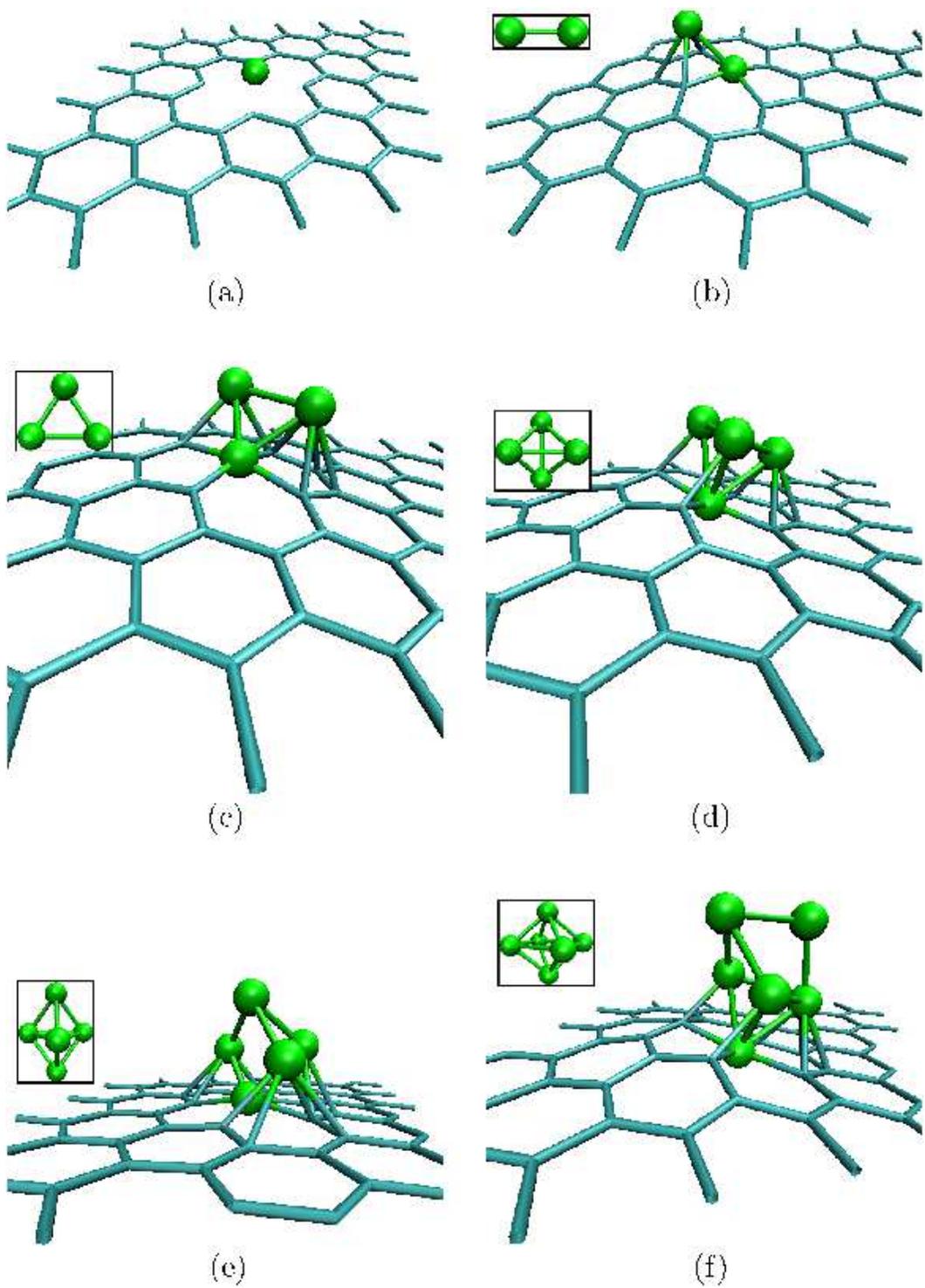}
\end{center}
\caption{Ground state geometries of Fe$_n$ ($n \le 6$) clusters on the
graphene sheet with a divacancy. Insets show the
geometries of corresponding clusters in free space.}
\label{fig:allgeom}
\end{figure*}

We begin our discussion by presenting the ground state geometries of the systems
studied. Our aim is to find out the possible evolution of the geometries of
clusters on the divacancy.  \ref{fig:allgeom} depicts all the geometries of
clusters on a divacancy, starting from Fe$_1$ to Fe$_6$. Also shown in the
insets are the geometries of free clusters.

As reported earlier,\cite{boukhvalov:09} the single iron atom diffuses almost to
the interior of the vacancy as seen in \ref{fig:allgeom}(a).  The four Fe-C
bonds are identical with the bond lengths of 1.97 \AA. It is interesting to note
that not only for a single atom but in general, the underlying vacancy does not
undergo any significant structural rearrangement.  This is in contrast with the
vacancy in absence of iron atoms where the carbon atoms are known to move close
to each other to form in-plane $\sigma$ bonds, resulting in a 5-8-5 ringed
structure \cite{coleman}. This indicates that even a single iron atom can be
used to maintain the structure of underlying lattice despite the presence of the
vacancy. 

As the size of the cluster increases some trend is discernible.  In all the
cases we note that one of the carbon atom is at the center of the vacancy. In
other words the structures evolve as a gradual addition of atoms on the geometry
of single iron system. 

Iron dimer has also been studied earlier \cite{boukhvalov:09} and our results
are in agreement with theirs. As seen from \ref{fig:allgeom}(b), the geometry is
obtained by adding an atom to the optimized geometry of Fe$_1$. The dimer bond
length is 2.20 {\AA} which is about 10\% enhanced over the bond length found in
a free cluster (2.02 {\AA}).  We recall that a single iron atom is just out of
plane, but forming four bonds with nearest neighbour carbon atoms, thus
maintaining the 5-8-5 ring structure.  Thus after one atom is accommodated in
this lattice, there is no more scope for any additional Fe atoms. Hence the
single atom remains with lattice plane and the additional Fe atoms are out of
the plane with the first atom as an apex .

It is worth mentioning that this system has a few isomers. The closest one is
the geometry where the dimer resides parallel to the plane of the graphene and
none of the iron atoms goes inside the vacancy.  The energy difference between
this structure and ground state is $\sim$ 0.3 eV.  This isomer is important
because  all  the larger clusters ($n >2$)  have at least one isomer with the
basic unit as a dimer parallel to the graphene plane.  The second isomer is
obtained by changing the orientation of the dimer  with respect to the graphene
plane. In this case the energy difference is about 0.4 eV.     

It is interesting to point out that, in the ground state, one of the atoms is
outside the vacancy which is  in contrast with a nitrogen dimer
\cite{biplabprb09} placed on a divacancy center.  In that case both the nitrogen
atoms of the dimer become a part of the graphene lattice by occupying the vacant
places of carbon atoms thereby completely healing the topological disorder.

Clusters evolve systematically as we go on adding an atom on Fe dimer. Fe$_3$
forms a triangle which is tilted with respect to vertical axis and none of the
sides are identical to each other.  The bond lengths are 2.22 {\AA}, 2.11 {\AA}
and 3.05 {\AA}.   The tilt occurs due to the optimization of the interactions
with underlying carbon atoms. Also, as the nearest carbon neighbours are not
same the sides of the triangle are not identical. 

When we add the fourth iron atom to the system, the triangle aligns vertically
and the fourth adatom positions from a side to form a distorted prism. Clearly
the distortion is brought about by the carbon lattice. As can be inferred from
the figure, the addition of the fifth atom is rather straightforward change from
prism to pyramidal structure with the added atom going on top. The sixth atom
distorts this pyramid and forms a complex structure as shown in
~\ref{fig:allgeom}.  Although distorted, the four-atom structure can still be
seen  near the graphene plane.  As we shall see later, the prism formed by four
atoms  is a very stable system and we believe that in more complex clusters with
large $n$ this prism may serve as the building block.  

Clearly the geometries of iron clusters on graphene are remarkably different
than those studied in free space \cite{yu}. As said earlier the dimer bond
length is slightly reduced in presence of divacant graphene.  The trimer in the
free space is reported to be isosceles triangle \cite{yu,gustev}, while the
trimer on vacancy is a distorted isosceles triangle with the bond lengths
differing substantially from that in free space.  Fe$_4$ also forms a prism in
free space with the bond lengths ranging from 2.22-2.41 {\AA}.  However in our
case the bond length varies substantially from 1.6 {\AA} to 2.6 {\AA}. The
change in the trigonal bipyramid of isolated Fe$_5$ is seen in the vertical
four-atom plane where the bond lengths are increased with respect to those in
free space.  Fe$_6$ undergoes a substantial change from octahedron to a more
complex structure seen in \ref{fig:allgeom}(f).

\subsection{Energetics and Magnetic structure}

Iron in a body centered cubic structure is known to be ferromagnetic in its bulk
phase as well as in a cluster form.  It is thus interesting to see the nature of
magnetism in the studied systems.  It is also helpful to discuss the energetics
with the same.  

\begin{figure}
\begin{center}
\includegraphics[width=10cm]{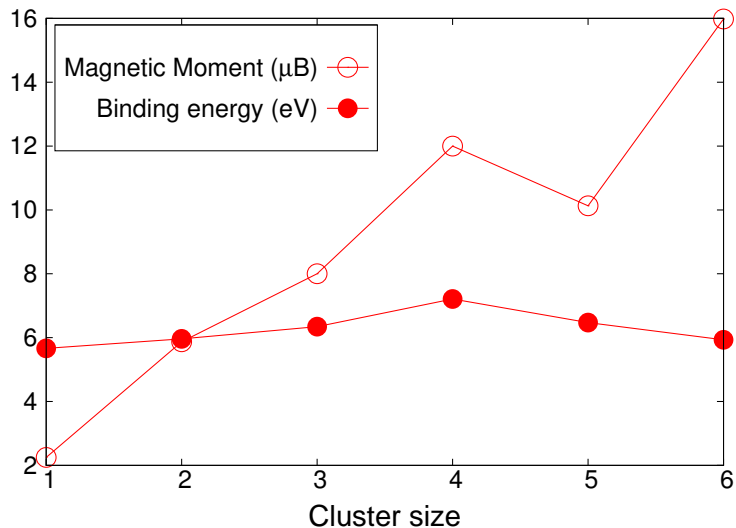}
\end{center}
\caption{Binding energies and total magnetic moments as a function of cluster size
(Fe$_n$).}
\label{fig:BeMu}
\end{figure}
Before we proceed we define the binding energy $\Delta_E$ of the iron cluster in presence of the divacancy as:
\begin{equation}
\Delta_E =  (E_{GD} + E_{Fe}) -E_{GD+Fe} 
\label{eq:be}
\end{equation} 

where $E_{GD+Fe}$ is the total energy of the system ({\it i.e.} the cluster on
graphene with a divacancy), $E_{GD}$ is the total energy of the graphene with a
divacancy and $E_{Fe}$ is that of an isolated iron cluster. Higher binding
energy indicates a more stable system.

\ref{fig:BeMu} shows the binding energies and total magnetic moments for all the
systems. Two  observations can be made immediately. Firstly, the binding energy
slowly increases from one to four atoms-cluster and then slowly decreases.
Secondly, the plot of total magnetic moment steadily increases except for the
five-atom cluster.  Thus from the observation of binding energies, it is not
surprising that we see a four-atom prism staying almost intact in larger
clusters. It also indicates that the Fe$_4$ is the most stable structure on the
divacancy center in graphene, however there is no indication of Fe$_4$ being the
most stable in the free space.

As seen from \ref{fig:BeMu}, all the systems studied are effectively magnetic in
nature. Our analysis of local magnetic moment has revealed that in all the cases
none of the carbon atoms has any significant local moment. Almost all the
contribution towards total magnetic moment is from iron atoms. Unlike the
isolated clusters it is inappropriate to calculate the magnetic moment per atom
as the atoms in the cluster may not have strictly iron neighbours.   

\begin{figure}
  \begin{center}
    \includegraphics[width=8cm]{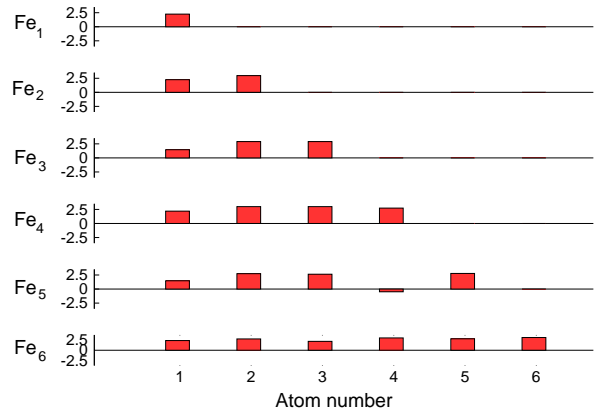}
  \end{center}
  \caption{Local magnetic moment on individual iron atoms for six clusters.}
  \label{fig:local}
\end{figure}

From \ref{fig:BeMu} it is clear that, up to the Fe$_4$ cluster,  as the number
of atoms increases, the magnetic moment also increases as the atoms are
ferromagnetically coupled to each other.  However as the cluster size increases,
the hybridization of Fe $d$ orbitals inside the clusters results in the lowering
of total magnetic moment. To ascertain that we also plot the local magnetic
moments on individual atoms. \ref{fig:local} shows the magnetic moments on
individual atoms for all the clusters. It can be deduced from the figure that as
the cluster increases in size the variation of the local moments becomes
non-monotonous (\ref{fig:BeMu}). Except Fe$_5$, all the other clusters studied
show aligned local moments on Fe atoms though they vary in size. In Fe$_5$
cluster, one of the iron atoms has the moment flipped with respect to the other
atoms. So, the average moment on iron in this cluster is smaller than the
others.

\begin{figure} \begin{center}
\includegraphics[width=8cm]{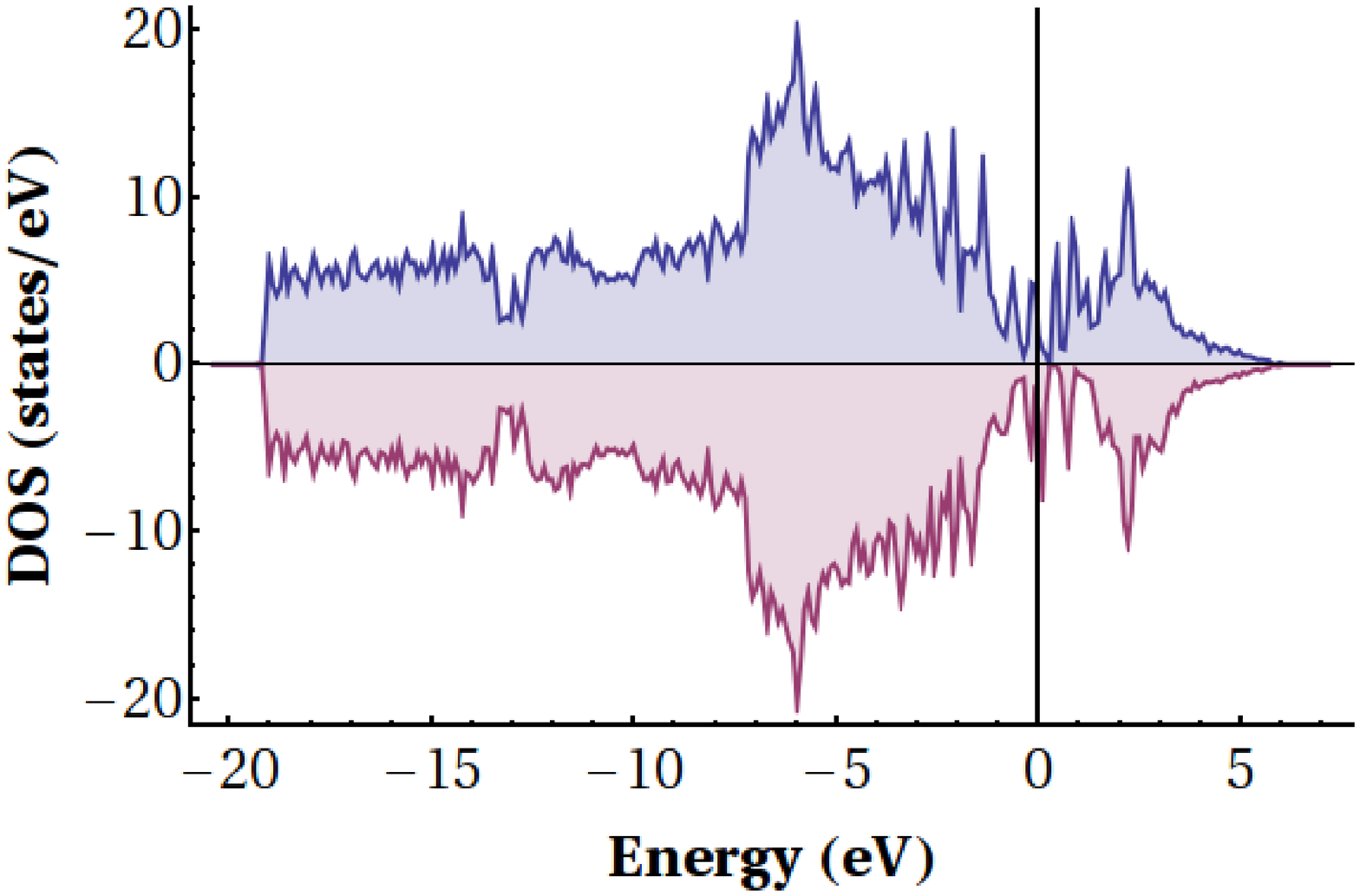}
\includegraphics[width=8cm]{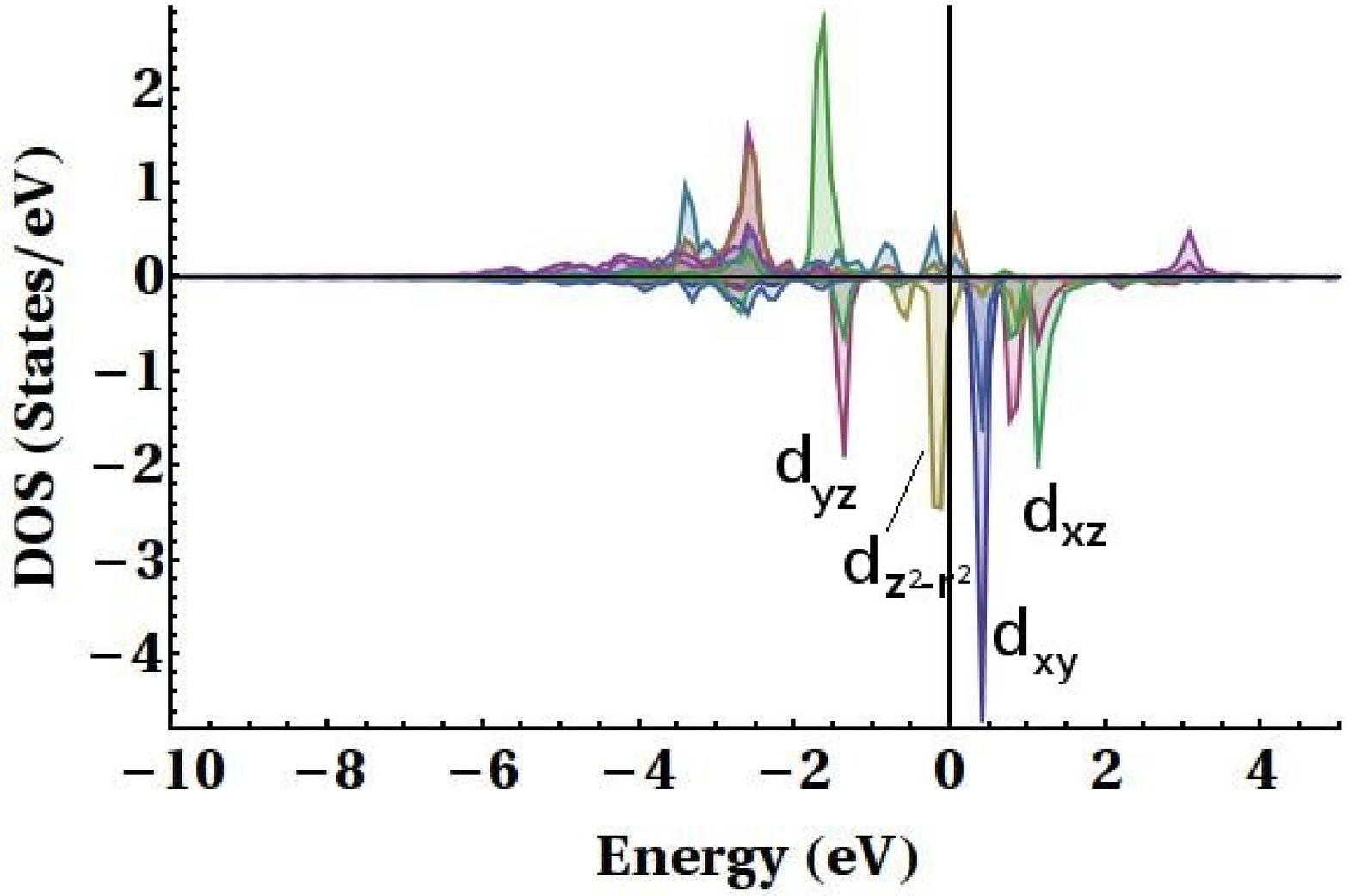}\\ {(a) \hskip 6cm (b)}
\end{center} \caption{ (a) Total DOS and (b) the local DOS of an iron atom on the divacancy. 
Only the $d$-components of LDOS are shown.
The upper and lower parts of the graph depict the spin-up and spin-down
components respectively.  Iron clearly induces electronic states on the Fermi level
which is marked by a vertical line.} \label{fig:fe1dos} \end{figure}

Naturally in the presence of of iron clusters the density of states (DOS) of the
graphene lattice undergoes a substantial change. Particularly interesting is the
fact that the contributions from the iron atoms occur directly at the Fermi
level.  \ref{fig:fe1dos} (a) shows the  total DOS in the presence of a single
iron atom while \ref{fig:fe1dos} (b) shows the local DOS (LDOS) on an iron atom.
LDOS is resolved into its angular components. It should be mentioned that the
single atom case is the simplest of all the cases studied however the features
seen here are similar for all the clusters. So, we present only the DOSs for a
single iron atom.

Our  angular momentum based analysis of LDOS on iron atom (\ref{fig:fe1dos} (b))
shows that the contribution from the components except $d$ is negligible. Also,
the characteristics of the electronic structure are very different from that of
pure graphene. The presence of midgap states of $p$ orbital character in
presence of a divacancy was discussed before \cite{coleman, jafri}.  Here, these
states are mainly of $d$ character. So, the characteristics of transport
properties are expected to be different.  As seen from the figure that the
contribution from the $d$ orbital is concentrated in the energy interval of
about 6 eV downwards from the Fermi energy. More interestingly, the contribution
from spin-up and spin-down channels of iron atom is substantially different. The
contribution near the Fermi level arising from the orbital with
$d_{z^{2}-r^{2}}$ character increases, maintaining the difference in the spin
channels as the cluster size increases. Such features are of particular interest
in spintronics applications where the spin-dependent transport properties are
discussed heavily.   

\begin{figure}
\begin{center}
\includegraphics[width=8cm]{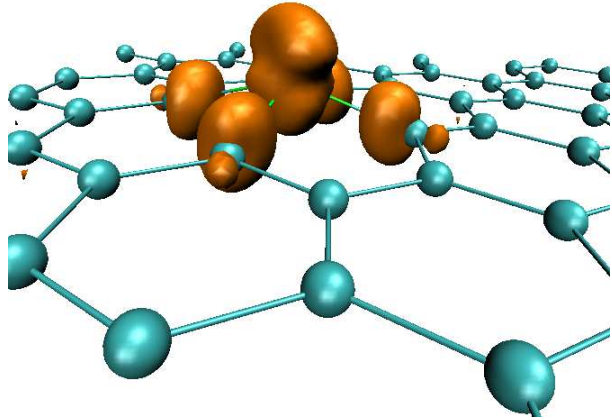}
\end{center}
\caption{Charge density isosurface of a state near the Fermi level. The system consists of
a single iron atom on which a $d$-like state is clearly seen and $p$-states on nearby
carbon atoms are heavily distorted. Isosurface is shown at one tenth of its
maximum value. \label{fig:fermi}}
\end{figure}

Thus it is clear that iron atoms induce a substantial number of states on the
Fermi level and it is seen that those are mainly $d$-like. This becomes further
evident from \ref{fig:fermi} which shows the charge density isosurface for a
state close to Fermi energy for a single iron atom. This feature is general and
is present in all the clusters.  The state represents a complex of $d$ states on
central iron atom and heavily distorted $p$ orbitals on neighbouring carbon
atoms.  Interestingly it is only the nearest carbon atoms those contribute
towards the charge density. It is the interaction between the four carbon atoms
with the iron cluster which is responsible for the strong binding.

\section{Conclusions}

We have systematically investigated the geometries and electronic structures of
Fe$_n$ ($n \le 6$) clusters on a graphene sheet with a divacancy using DFT.
When a divacancy is created in graphene, the underlying structure undergoes
structural rearrangements to form a 5-8-5 ringed structure.  The presence of an
iron cluster, however prevents the formation of 5-8-5 ring and maintains the
pristine hexagonal structure.  The geometries of iron clusters also undergo
significant changes due to the interaction between their $d$ orbitals with $p$
orbitals of the neighbouring carbon atoms. Individual atom in the iron cluster
possesses  different magnetic moment and in general the atom closest to carbon
matrix has the lowest magnetic moment. However as the cluster size increases,
more complex patterns of magnetic moments emerge. The iron clusters in free
space has an average moment $\sim$ 3 $\mu_B$, however in presence of divacant
graphene the average moment is reduced to $\sim$ 2.5 $\mu_B$. The iron cluster
also has an important contribution in DOS, as most of the $d$ states of iron
contribute within $\pm$ 3 eV around the Fermi level. Our spin polarized
calculations also reveal that the contribution of up and down channels are not
identical making the system a candidate for spintronics materials.

\begin{acknowledgement}
B.S.P. would like to acknowledge  CSIR, Govt. of India  (No:
9/137(0458)/2008-EMR-I). B.S. gratefully acknowledges VR/SIDA funded Asian-Swedish Research Link Program, Carl Tryggers Foundation, G\"{o}ran Gustafssons Foundation and STINT  for financial support. We are grateful to HPC2N and UPPMAX under Swedish National Infrastructure for Computing (SNIC) for providing computing facility. Some of the figures are generated by using VMD software \cite{vmd}.  
\end{acknowledgement}

 \bibliography{biblio}

\end{document}